\documentclass{iopart}
\usepackage{bbm,amssymb,amsbsy,times,pifont}
\usepackage[dvips]{graphicx}
\usepackage{hyperref,graphicx,subfigure}
\usepackage{color,amsfonts,amssymb,graphics,epsfig}

\newcommand{\R}{\mathbbm{R}}

\newcommand{\id}{\mathbbm{1}}

\renewcommand{\tr}{{\rm Tr}\,}
\renewcommand{\det}{{\rm Det}\,}
\newcommand{\gr}[1]{\boldsymbol{#1}}
\newcommand{\be}{\begin{equation}}
\newcommand{\ee}{\end{equation}}
\newcommand{\bea}{\begin{eqnarray}}
\newcommand{\eea}{\end{eqnarray}}
\newcommand{\ket}[1]{|#1\rangle}

\newcommand{\sig}{\gr{\sigma}}

\newcommand{\eq}[1]{Eq.~(\ref{#1})}

\begin{document}
\title{Canonical and micro-canonical typical entanglement of continuous variable systems}
\author{A. Serafini, O. C. O. Dahlsten, D. Gross, and M. B. Plenio}
\address{Institute for Mathematical Sciences, 53 Prince's Gate, Imperial College London, London SW7 2PG, UK
and QOLS, Blackett Laboratory, Imperial College London, London SW7 2BW, UK}
\begin{abstract}
{We present a framework, compliant with the general canonical principle of statistical mechanics, 
to define measures on the set of pure Gaussian states of continuous variable systems. 
Within such a framework, we define two specific measures, referred to as `micro-canonical' and `canonical', 
and apply them to study systematically the statistical properties of the bipartite entanglement of 
$n$-mode pure Gaussian states (as quantified by the entropy of a subsystem). 
We rigorously prove the ``concentration of measure'' around a finite average, 
occurring for the entanglement 
in the thermodynamical limit in both the canonical and the micro-canonical approach. 
For finite $n$, we determine analytically the average and standard deviation of the entanglement 
(as quantified by the reduced purity)
between one mode and all the other modes. Furthermore, we numerically investigate more general situations,
clearly showing that the onset of the concentration of measure already occurs at relatively small $n$.
}
\end{abstract}
\maketitle

\section{Typical entanglement in quantum information theory}
Due to the exponentially increasing complexity 
of the Hilbert spaces of multiple constituents, a complete theoretical characterisation 
of the entanglement of general quantum systems of many particles 
turns out to be a daunting task \cite{Plenio V 07}.
A viable approach towards such a characterisation consists in focusing on the 
``typical'', {\em statistical} properties of the quantum correlations of multipartite systems, 
when the states of the system are assumed to be distributed according to a particular `measure'. 
This strategy is firstly aimed at simplifying the problem at hand by restricting attention on the 
typical (and thus, in a sense to be precisely specified in the following, ``overwhelmingly likely'') 
features of the entanglement of a system whose state is apt to be described by the chosen measures.
Furthermore, this kind of analysis is able to shed light on the general properties of the 
entanglement of physical systems.

For finite dimensional quantum systems, a natural, `uniform' measure on pure states emerges from 
the ``Haar'' measure of the unitary group
({\em i.e.}, from the left- and right-invariant measure under application of any unitary transformation), 
whose elements allow to retrieve any state when applied to another given starting pure state.
On such grounds, a well defined typical entanglement of finite dimensional systems can be addressed 
and analysed.
Original studies in this direction were undertaken well before the development
of the formal theory of entanglement developed in quantum information science \cite{Plenio V 07}: 
In 1978, Lubkin considered the expected entropy
of a subsystem when picking pure quantum states at random from the uniform
measure\cite{lubkin}. 
Let us recall that the von Neumann entropy $S=-tr\rho \ln{\rho}$ of a subsystem in state $\rho$ 
properly quantifies, for globally pure states, 
the entanglement between the subsystem and the remainder of the system. 
Lubkin showed that one expects this quantity to be nearly maximal. 
Pagels and Lloyd arrived later at the same qualitative conclusions, following 
an independent line of thought \cite{pagels}. 
Their work was expanded by Page, who conjectured an exact formula for the average 
entropy of a subsystem $S_{m,n}$ \cite{page}, reading
\begin{equation}
\label{eq:page}
S_{m,n}=\sum_{k=n+1}^{mn}\frac{1}{k}-\frac{m-1}{2n},
\end{equation} 
for a quantum system of Hilbert space dimension $mn$ in a random pure state, 
and a subsystem of dimension $m\leq n$. 
This relation was later proven by Foong and Kanno \cite{foong}.

This general line of enquiry was revisited and considerably extended in the setting of quantum information
theory by Hayden, Leung and Winter in Ref.~\cite{hayden}, where they extensively
studied the `concentration of measure' around the average of the entanglement probability distribution
with increasing $n$. 
They also pointed out that this study may provide a way of simplifying the theory of entanglement 
which contains a plethora of locally inequivalent classes. 
In other words, 
as already mentioned,  
restricting statements to the ``typical entanglement'' allows one to ignore several unessential complications.
Recent results on the physical interpretation of Page's conjecture can be found in
Refs.~\cite{oliveira, dahlsten}, 
where it is proven that a circuit of elementary quantum gates on a quantum circuit  
is expected to maximally entangle the state to a fixed arbitrary accuracy, 
within a number of gates that grows only polynomially in the number of qubits of the register.   

A further simplification in the analysis of the entanglement 
can be achieved by considering the typical entanglement of `particularly relevant' 
(according to the specific problem at hand) subsets of states. 
For example, it has been found that `stabilizer states' 
(a countable set of states playing a central role in quantum error correcting codes)
are also typically maximally entangled \cite{smith,dahlsten2}, similarly to the set of all states.
Such investigations are interesting {\em per se}, 
as they unveil the potential and limitations hidden in the adoption of restricted classes of states
and, furthermore, provide us with a more detailed understanding of the
entanglement properties of the total state space.

Here we expand these considerations into the realm of continuous variables,

{\em i.e.} of quantum systems described by pairs of canonically conjugated
observables with continuous spectra. Such systems, ranging from motional degrees of freedom

of particles in first quantisation to bosonic fields in second quantisation, are ubiquitous to all areas of quantum physics, being prominent
in quantum optics (as they embody the light field in second quantisation), atomic physics (notably, in the description of atomic ensembles), quantum field theory (as they encompass any bosonic field), in addition to their
crucial role in molecular and atomic physics.  
 
Quantum systems described by operators with continuous spectra live in infinite dimensional 
Hilbert spaces. 
Therefore, a first naive try in this direction
could be to take the infinite dimensional limit of \eq{eq:page}. 
Page showed that \eq{eq:page} implies that 
$S_{m,n}\simeq \ln{m} -\frac{m}{2n}$ for $1\ll m \leq n$. Under that restriction, and noting that
the maximal entanglement is $\ln{m}$, we can then make the observation that
the ratio of the entanglement average to the maximum tends to unity as $n\longrightarrow\infty$,
but that the two quantities both diverge logarithmically. 
From this perspective, 
the entanglement could be said to be typically infinite in the continuous variable setting.
Even though mathematically reasonable (in the limit's sense), this statement is definitely  
questionable, physically and practically. In fact, in any practical situation, one will deal with a 
finite total energy or with finite ``temperatures'' (both quantities will be defined precisely
in our treatment), whereas infinitely entangled states require an infinite energy to be created. 
In this paper, we shall restrict our attention on pure Gaussian states whose 
typical entanglement we shall study under two different measures, both inspired by arguments 
of thermodynamical natures, but apt to describe different situations.
The previously mentioned divergences, still potentially emerging in the Gaussian setting, 
will be tamed by introducing proper prescriptions which will generally affect the energy of the system,
by imposing either a sharp upper bound or an exponentially decaying distribution of energies
(in this respect, 
see also Ref.~\cite{jpa} for a detailed discussion of the impact of similar constraints on entanglement 
measures).
We will show that such prescriptions induce the occurrence of a {\em finite} typical entanglement 
in the limit of an infinite number of total constituents. 
Even for a finite number of total degrees of freedom and finite upper bound to the energy 
-- entailing the existence of a finite maximal entaglement -- 
the typical entanglement concentrates around a value which is well distant 
from the allowed maximum.

Notice that Gaussian states are certainly the most prominent class of states not only in quantum information 
with continuous variables, but also, more broadly, in quantum optics, 
as they can be generated and manipulated with relative ease (even as highly 
entangled states \cite{generation1,generation2,generation3}), 
can be used for the implementation 
of quantum communication and information protocols \cite{braunsteinvanloock} 
and serve as a powerful testing ground 
for the theoretical characterisation of entanglement properties \cite{martinjens}.

The micro-canonical measure, which we will apply here to the study of the 
typical entanglement, has been already employed in the analysis of the quantum teleportation 
of Gaussian states with generic second moments \cite{nosotros}.  
In particular, the micro-canonical 
average quantum fidelity and a corresponding ``classical threshold'', 
have been evaluated for the teleportation of states with null first moments and arbitrary second moments 
under the standard continuous variable teleportation protocol (see, {\em e.g.}, \cite{braunsteinvanloock} 
for a description of the scheme).

Let us finally mention that 
a definition of micro-canonical average entanglement has been very recently addressed for 
finite dimensional systems as well \cite{verhulst}, 
with a major emphasis on the possibility of reducing 
time-averages to ensemble-averages.

This paper is organised as follows. 
In Section \ref{techintro} we review some preliminary facts about Gaussian states and 
set the notation. In Section \ref{measures} we review the definition of the micro-canonical measure 
on pure Gaussian states, already introduced in \cite{nosotros}, completing 
our previous analysis with the inclusion of comments and mathematical details 
previously omitted, and we extend the existing framework to encompass a `canonical' measure as well.
Section \ref{sec:conce} contains a rigorous proof of the `concentration of measure', common to 
the two measures introduced here, {\em i.e.}~of the fact that the entanglement probability distribution concentrates 
in the thermodynamical limit, around a finite `thermal' average, away from the allowed maximum. 
Even though this result had been anticipated in Ref.~\cite{nosotros}, this proof is original and adds 
further insight into the matter. Section \ref{study} presents a detailed study about 
the typical entanglement of pure Gaussian states with finite number of degrees of freedom, 
where both analytical findings and numerical evidences are reported.
Conclusions and outlook are found in Section \ref{outlook}. Three appendices complement the work, 
one of which (\ref{mathematicalize}) 
contains the most technical steps needed to prove the concentration of measure, 
while in \ref{bch} a specific Baker-Campbell-Hausdorff relation is derived, and 
in \ref{maximal} a derivation of the expression of the maximal entanglement for given energy 
is presented.

\section{Preliminaries}\label{techintro}

We consider bosonic continuous variable (CV) quantum mechanical systems 
described by $n$ pairs of canonically conjugated 
operators $\{\hat x_j,\hat p_j\}$ 
with continuous spectra, like motional degrees of freedom of particles in first quantisation
or bosonic field operators in second quantisation.
Grouping the canonical operators together in the 
vector $\hat R=(\hat x_1,\ldots,\hat x_n,\hat p_1,\ldots,\hat p_n)^{\sf T}$ allows to 
express the canonical commutation relations (CCR) as
$
[\hat R_j,\hat R_k] = 2i\,\Omega_{jk} 
$, 
where the `symplectic form' $\Omega$ is defined as 
$$
{\Omega} = \left(\begin{array}{cc}
0_n & \id_n \\
-\id_n & 0_n
\end{array}\right) \, ,
$$
$0_n$ and $\id_n$ standing for the null and identity matrix in dimension $n$.

Any state of an $n$-mode CV system is described by a positive, trace-class 
operator $\varrho$. 
For any state $\varrho$,
let us define the $2n$-dimensional vector of the expectation 
values (``first moments'') of the canonical operators $R$ (with entries $R_j$) as 
$$
R_{j} \equiv \tr{[\hat R_j \varrho]}
$$
and the $2n\times 2n$ matrix of second moments, 
or ``covariance matrix'', 
${\gr\sigma}$ (with entries $\sigma_{i,j}$) as
$${\sigma}_{i,j} \equiv \tr{[\{\hat R_i , \hat R_j\} \varrho]}/2
-\tr{[\hat R_i \varrho]}\tr{[\hat R_j \varrho]} \, .$$

Also, throughout the paper, we will refer to the `energy' of a state $\varrho$ as to the 
expectation value of the operator $\hat H_0 = \sum_{j=1}^{n} (\hat{x}_j^2+\hat{p}_j^2)$.
This definition corresponds to the energy of a free electromagnetic field in the 
optical scenario (and to decoupled oscillators in the general case). 
In our convention, as determined by the factor $2$ appearing in the CCR,
the vacuum of a single mode has covariance matrix equal to the identity (thus simplifying 
significantly several expressions), with energy $2$ 
(the adopted energy unit is $\hbar\omega/4$ for a mode of frequency $\omega$).
The energy is determined by first and second moments according to 
$$
\tr{(\varrho\hat{H}_0)} = \tr(\sig) + \|R\|^2  \; ,
$$
where $\|R\|$ is the usual euclidean norm of the vector $R$.

Gaussian states are defined as the states with Gaussian characteristic functions 
and quasi-probability distributions, defined over a phase space analogous to that of 
classical Hamiltonian dynamics.
As well known, a pure state $\ket{\psi}_{G}$ is Gaussian if and only if it can be 
obtained by transforming 
the vacuum $\ket{0}$ under an operation generated by 
a polynomial of the second order in the canonical operators. 
In formulae (up to a negligible global phase factor):
\be
\ket{\psi}_{G} = \hat{G}_{A,b} \ket{0} \equiv 
\,{\rm e}^{i(\hat{R}^{\sf T} A \hat{R}+\hat{R}^{\sf T}b)} \ket{0} \; , \label{pures}
\ee
where $A$ and $b$ are, respectively, a real $2n\times2n$ matrix and a real $2n$-dimensional vector.
Because of the CCR and of the unitarity of 
$\hat{G}_{A,b}\equiv\,{\rm e}^{i(\hat{R}^{\sf T} A \hat{R}+\hat{R}^{\sf T}b)}$, 
$A$ can be chosen symmetric, without loss of generality, while $b$ is a generic real vector. 
$A$ and $b$ determine the CM and the second moments of the Gaussian state, 
thus completely determining it.

The unitary operator can always be rewritten as (see \ref{bch})
\be
\hat{G}_{A,b} = \, {\rm e}^{i(\hat{R}^{\sf T} A \hat{R})} 
\, {\rm e}^{i(\hat{R}^{\sf T} M b)} \, ,
\label{pures2}
\ee
for some matrix $M$ (as shown in \ref{bch}, $M=\Omega A^{-1}(\id_{2n}-\,{\rm e}^{4A\Omega})/4$ 
for invertible $A$'s). 
First and second order operations can thus be generally `decoupled'.

First order operations correspond to local displacements in phase space. While 
such operations do not affect local entropies (and thus the entanglement) of multipartite states, 
they do affect the energy of the states, which will play a central role in what follows.
Moreover, let us notice that 
the group of these transformations is non-compact, 
being isomorphic to the abelian $\R^{2n}$ under the addition composition rule.
In the following, we will show how the first moments can be consistently 
incorporated in the presented framework. 
However, because their inclusion in the study of the 
statistical properties of the entanglement is just a technicality (adding no significant insight), 
we will set them to zero in the investigations to come.

As for second order transformations, determined by the matrix $A$, they can be conveniently mapped 
into the group $Sp_{2n,\R}$ of real {\em symplectic transformations}, acting linearly in phase space 
(as second order transformations acting on the Hilbert space
make up the multi-valued {\em metaplectic} representation of the symplectic group \cite{folland}).
Recall that a matrix $S$ belongs to the symplectic group $Sp_{2n,\R}$ 
if and only if it preserves the antisymmetric form $\Omega$: 
$S\in SL({2n,\R}) \, :\;
S\in Sp_{2n,\R} \Leftrightarrow S^{\sf T} \Omega S = \Omega$. 
Let us also recall that a symplectic transformation $S$ 
acts by congruence on a covariance matrix $\sig$: $\sig\mapsto S^{\sf T}\sig S$.
Of course, symplectic transformations can in general affect both the entanglement and the energy 
of a state. 
The algebra of generators of the symplectic group is comprised of all the matrices that can be written as 
$\Omega J$, where $J$ is some $2n\times2n$ symmetric matrix \cite{pramana}
(in this notation, generators are not complexified, so that $S=\,{\rm e}^{\Omega J}$). 
Such generators do not have a definite symmetry
({\em i.e.}, they are not necessarily symmetric or antisymmetric). 
Choosing a basis of the algebra such that each generator of the basis is either symmetric or antisymmetric, 
allows one to distinguish between a {\em compact} subgroup $K(n)=Sp_{2n,\R}\cap SO(2n)$ 
(spawned by antisymmetric generators) and a {\em non-compact} subgroup (arising from skew-symmetric 
generators). Notice also that, since compact transformations are indeed orthogonal, they do not 
affect the energy of the states they act upon (explicitly, $\tr{\sig}$ and $\|R\|^2$ 
are both invariant under phase space ``rotations''). 

Remarkably, the subgroup $K(n)$ is isomorphic to $U(n)$. 
Because this fact will be exploited throughout the whole work, we shall sketch its proof here.
Let us define the transformation $O$ by
$$
O = \left(\begin{array}{cc}
X & Y \\
W & Z
\end{array}\right) \; , 
$$
where $X$, $Y$, $W$ and $Z$ are $n\times n$ real matrices. 
It is straightforward to show that this transformation 
is symplectic and orthogonal if and only if $Z=X$, $W=-Y$, 
$X^{\sf T}X+Y^{\sf T}Y = \id$ and $X^{\sf T}Y-Y^{\sf T}X = 0$, so that 
\be
O = \left(\begin{array}{cc}
X & Y \\
-Y & X
\end{array}\right) \; . \label{isu}
\ee
Now, let $U=X+iY$ be a matrix with real part $X$ and imaginary part $Y$. 
The unitarity condition on $U$ corresponds exactly to the previous two conditions on 
$X$ and $Y$, thus demonstrating the existence of a bijective 
mapping from $U(n)$ to $K(n)$. The preservation of the composition rule can be 
straightforwardly checked out.
Incidentally, this isomorphism implies that $K(n)$ has $n^2$ independent parameters.

Let us rephrase \eq{pures} to give a transparent parametrisation of pure Gaussian states
in phase space terms, by considering the action of first and second order operations on the 
covariance matrix and on the first moments:
\be
\sig = S^{\sf T} S   \, ,\quad {\rm with} \quad S\in Sp_{2n,\R} 
\, , \quad R \in \R^{2n} \,  \label{pureps}
\ee
(recall that, in our units, the covariance matrix of the vacuum is the identity).
Indeed, because of the peculiar nature of their characteristic functions, Gaussian states 
are completely determined by first and second moments of the canonical operators.

More generally, let us also recall that the CM ${\gr \Sigma}$ 
of any, pure or mixed, Gaussian state can be written as 
\be
\gr{\Sigma} = S^{\sf T} \gr{\nu} S \; , \label{williamson}
\ee
where $S\in Sp_{2n,\R}$ and $\gr{\nu}=\,{\rm diag}(\nu_1,\ldots,\nu_n,\nu_1,\ldots,\nu_n)$ 
is a diagonal matrix with double-valued eigenvalues called the ``Williamson normal form'' 
of $\gr{\Sigma}$ \cite{williamson36,simon99} (corresponding to the normal-modes decomposition 
of positive definite quadratic Hamiltonians). 
The real quantities $\{\nu_j\}$ 
are referred to as the `symplectic eigenvalues' of $\gr{\Sigma}$ and can be computed as the 
eigenvalues of the matrix $|i\Omega\gr{\Sigma}|$.
The symplectic eigenvalues 
hold all the information about the entropic quantities of the Gaussian state in question. 
In particular, the `purity' $\mu\equiv \tr\varrho^2$ of the Gaussian state $\varrho$ 
with CM $\gr{\Sigma}$ is determined as 
\be
\mu=1/\prod_{j=1}^{n}\nu_j=1/\sqrt{\det\gr{\Sigma}}\,, , \label{purity}
\ee
while the von Neumann entropy $S\equiv-\tr(\varrho\ln\varrho)$ reads 
\be
S=\sum_{j=1}^{n}h(\nu_j) \, , \label{vneu}
\ee
with the `entropic function' $h(x)$ given by
\be
h(x) = \frac{x+1}{2}\log_{2}(\frac{x+1}{2}) - \frac{x-1}{2}\log_{2}(\frac{x-1}{2}) \, . \label{entrofunc}
\ee
Being the eigenvalues of $|i\Omega\gr{\Sigma}|$, the symplectic eigenvalues are 
continuously determined 
by `symplectic invariants' ({\em i.e.} by quantities depending on the entries of the CM invariant 
under symplectic transformations), 
defined as the coefficients of the characteristic polynomial of such a matrix \cite{serafozzi06,serafozzijosab}).
This observation will be useful later on.

\section{Measures on the set of pure Gaussian states}\label{measures}

The present section is devoted to the definition of consistent measures 
on the set of pure Gaussian states, introducing a broad framework 
motivated by fundamental statistical arguments. 
We will review the construction of the `micro-canonical measure', 
already introduced in Ref.~\cite{nosotros}, complementing such earlier studies with 
discussions and mathematical details. 
Furthermore, within the same general framework, 
we will present a novel, ``canonical'' measure, thus extending our previous treatment.

A `natural' measure to pick would be one invariant under the action of the operations which generate 
the set of states we are focusing on. In the previous section, we have analysed such a set of operations
for pure Gaussian states, showing that it amounts to symplectic operations and displacements.
One would thus be tempted to adopt the left- and right- invariant measure ({\em i.e.}, the {\em Haar} measure)
over such groups.  
Unfortunately, because the symplectic group is non compact, the existence of a Haar measure 
on the whole group [from which a measure for pure Gaussian states could be derived via \eq{pureps}] 
is not guaranteed. Notably, even if such a measure could be constructed, it would 
not be normalisable, giving rise to distributions with unbounded statistical moments.
Moreover, some prescription has obviously to be introduced also to handle the first moments 
which are, in general, free to vary in the non-compact $R^{2n}$ (notice that the Euclidean volume
is obviously invariant under left translations but 
is not a proper measure in the space of first moments 
because it is not normalisable, due to the non-compactness of $R^{2n}$). 

To cope with such difficulties we will introduce, in analogy with statistical mechanical treatments, 
assumptions on the energy of the states under examination, which will constitute our `privileged' 
physical observable (in a sense to be elucidated in the following).
A proper structure to introduce a measure is inspired by 
a well known decomposition of an arbitrary symplectic transformation $S$: 
\begin{equation}
S = O' Z O ,\label{euler1}
\end{equation}
where $O, O' \in K(n)= Sp(2n,\R)\cap SO(2n)$ are orthogonal
symplectic transformations, while
\be
Z=Z'\oplus Z'^{-1} \; , \label{defZ}
\ee
where $Z'$ is a diagonal matrix with eigenvalues $z_{j}\ge 1$ $\forall$ $j$.
The set of such $Z$'s forms a non-compact subgroup of $Sp_{2n,\R}$ 
(corresponding to local squeezings), which will be denoted by $Z(n)$. 
The virtue of such a decomposition, known as ``Euler'' (or ``Bloch-Messiah'' \cite{braun05}) decomposition,
is immediately apparent, as it allows one to distinguish between the degrees 
of freedom of the compact subgroup (essentially `angles', ranging from $0$ to $2\pi$,
which moreover do not affect the energy) 
and the degrees of freedom $z_j$'s with non-compact domain. 
In particular, applying Euler decomposition to \eq{pureps} leads to
\be
\sig = O^{\sf T} Z^2 O \; . \label{purecm}
\ee 
Due to the rotational invariance of the vacuum in phase space, the number of free 
parameters of a pure Gaussian state of an $n$-mode system is thus $n^2 + 3n$
(taking the $2n$ independent first moments into account). 

Quite naturally, we shall assume 
the $n^2$ parameters of the transformation $O$ to be distributed according 
to the Haar measure of the compact group $K(n)$, carried over from $U(n)$ 
through the isomorphism described by \eq{isu}. 
The set of such parameters will be compactly referred to as $\vartheta$, 
while the corresponding Haar measure will be denoted by ${\rm d}\mu_{H} (\vartheta)$.

We have thus identified the variables which parametrise an arbitrary pure Gaussian state
and imposed a distribution on a subset of them.
A `natural' measure has yet to emerge for the non-compact variables $\{z_j\}$ and $\{R_j\}$.
In order to further constrain the choice of measures on such variables we shall invoke now a fundamental
statistical argument. In their kinematical approach to statistical mechanics \cite{popescu05}, Popescu, Short and 
Winter introduced a general principle, 
which they refer to as {\em general canonical principle}, stating that
\begin{quote}
``Given a sufficiently small subsystem of the universe, almost every pure state of the universe 
is such that the subsystem is approximately in the `canonical state' $\varrho_{c}$.''
\end{quote}
The `canonical state' $\varrho_c$ is, in our case, the local reduction of
 the global state picked from a distribution of states with maximal entropy
 under the constraint of a maximal total energy $E$. That is, 
quite simply, a ``thermal state'', which is a Gaussian state with null first moments and CM $\sig_c=(1+T/2)\id$. 
Here the `temperature' $T$ is defined by passage to the ``thermodynamical limit'', that is  
for ${n\rightarrow\infty}$ and ${E\rightarrow\infty}$,   
$(E-2n)/n\rightarrow T$ (assuming $k_{B}=1$ for the Boltzmann constant). 
For ease of notation, in the following,
the symbol $\simeq$ will imply that the equality holds in the thermodynamical limit, 
{\em e.g.}:~$(E-2n)/n\simeq T$.
Notice that we have required here the introduction of a maximum energy $E$ or, alternatively, 
of a temperature $T$.
In point of fact, such requirements are necessary to handle the non-compact part of the symplectic group. 
Note also that, in principle, two options are open in this respect, as one can introduce either an upper bound
to the energy or a temperature: essentially, these two distinct options characterise the two distinct approaches 
(micro-canonical and canonical) which we will detail in the next section.

Because the canonical state $\varrho_c$ is Gaussian with vanishing first moments, the general canonical principle
can be fully incorporated into our restricted (Gaussian) setting.
As we have shown in Ref.~\cite{nosotros}, the compliance with the general canonical principle 
enforces a rather stringent restriction on the distribution of the non-compact variables. 
In particular, the general canonical principle is always satisfied if, 
{\em in the thermodynamical 
limit, such variables are independent and identically distributed (i.i.d)} \cite{gennote}.
To keep our exposition lighter and more readable,
we will not repeat here the technical derivation of this implication. 
Actually, it will be entirely subsumed, {\em a posteriori}, by the derivation 
of the ``concentration of measure'' in Sec.~\ref{sec:conce}.

Before moving on with the definition of specific measures, let us comment on the first moments $\{R\}$. 
The general canonical principle imposes that, in the thermodynamical limit, 
their density of probability $p'(R)$ tend to a $\delta$-distribution centered in $0$ 
(in fact, the canonical state has vanishing first moments). 
A suitable example is $p'(R)=(n\lambda/\pi)^n\exp(-n \lambda \|R\|^2)$ for some constant $\lambda$ 
(notice that $\|R\|^2$ is the first moments' contribution to the energy). 
Let us remark that this class of distributions encompasses the ones usually adopted 
for coherent states, in the computation of classical teleportation thresholds \cite{teleport1,teleport2}.
From now on, we will just set the first moments to zero: they can be coherently incorporated 
into our general picture following the recipe given above.

Let us now turn to second moments and sum up our line of thought so far:
inspired by mathematical considerations and guided by physical arguments,
we have defined a distribution for the ``compact'' degrees of freedom $\vartheta$ 
(essentially, their Haar measure) and specified a prescription for the non-compact parameters $z_{j}$'s.
Several choices are then possible, within this prescription, to deal with the variables $z_j$'s.
We will describe now in detail two of such choices, which we will then apply to study the 
typical entanglement.

\subsection{Micro-canonical measure}

As a first approach, we will introduce a micro-canonical measure 
on the class of $n$-mode pure Gaussian states with an energy upper bounded by $E$. 
Notice that such a restriction is essentially equivalent to {\em fixing} 
the total energy of $(n+1)$-mode states to $E+2$ 
(in fact, in the latter instance, the energy of the additional mode is not independent and merely 
`makes up' to reach the fixed total amount). Notably, the two approaches are obviously indistinguishable 
in the thermodynamical limit.

Notice that the parameters $z_j$, whose distribution is left to define,
determine the energy $E_j$ pertaining to each {\em decoupled} mode $j$
of the Euler decomposition, according to (see \eq{purecm})
\be
E_j = z_j^2 + \frac{1}{z_j^2} \; . \label{enej}
\ee
Here, 
we will assume a Lebesgue (`flat') measure for the local energies $E_j$'s 
(uniquely determining the squeezings $z_j$'s, 
as $z_j\ge 1$), inside the region $\Gamma_{E}=\{\gr{E} : |\gr{E}|\le E\}$ 
bounded by the linear hypersurface of total energy $E$ (here, ${\gr E}=(E_1,\ldots,E_n)$ 
denotes the vector of energies, with all positive entries, while $|\gr{E}|=\sum_{j=1}^{n}E_j$). 
More explicitly, denoting by ${\rm d}p_{mc}(\gr{E})$ the probability of the occurrence of the energies $\gr{E}$, 
one has 
\bea
{\rm d}p_{mc}(\gr{E}) &=& {\cal N} \,{\rm d}^n\gr{E} \equiv {\cal N}\,{\rm d}E_1\ldots{\rm d}E_n   \quad {\rm if} \quad \gr{E}\in \Gamma_{E}\, , \nonumber \\
{\rm d}p_{mc}(\gr{E}) &=& 0 \quad {\rm otherwise} \; ,
\eea
where ${\cal N}$ is a normalisation constant equal to the inverse of the volume of $\Gamma_{E}$.
Notice that such a flat distribution is the one maximising the entropy in the knowledge of the local energies 
of the decoupled modes. In this specific sense such variables have been privileged, on the basis
of both mathematical (the Euler decomposition) and physical (analogy with the micro-canonical ensemble) grounds.
Let us also mention that, as will become apparent in the next section,
employing the variables $\{E_{j}\}$ leads to a remarkable simplification of the expression 
of the averages over the Haar measure of the 
compact subgroup. 
While purely formal, this aspect
yields some significant insight into the privileged role of such variables
in characterising the statistical 
properties of physical quantities.

The micro-canonical average $Q_{mc}(E)$ over pure Gaussian states at 
maximal energy $E$ of 
the quantity $Q(\gr{E},\vartheta)$ determined by the second moments alone 
will thus be defined as 
\be
Q_{mc}(E) = {\cal N}\int {\rm d}\mu_{H}(\vartheta) \int_{\Gamma_{E}} {\rm d}\gr{E} \, Q(\gr{E},\vartheta) \; ,  \label{micro}
\ee
where the integration over the Haar measure is understood to be carried out over the whole compact domain 
of the variables $\vartheta$.
More explicitly, the integral over the energies $E$ can be recast as
\be
\int_{\Gamma_{E}} {\rm d}\gr{E} = {\cal N}
\int_{2}^{E-2(n-1)}{\rm d}E_1 
\ldots \int_{2}^{E-\sum_{j=1}^{n-1}E_j}{\rm d}E_n \label{recast}
\ee
(each energy is lower bounded by the vacuum energy, equal to $2$ in the convention adopted here),
determining the normalisation as ${\cal N}=n!/(E-2n)^{n}$ and 
leading to a marginal density of probability $P_{n}(E_j,E)$ for each of the energies $E_j$ given by
\be
P_{n} (E_j,E) = \frac{n}{E-2n} \left( 1- \frac{E_j-2}{E-2n} \right)^{n-1} \; . \label{marginal}
\ee
Clearly, the energies $E_{j}$ are not i.i.d.~for finite $n$.
However,
as is apparent from Eqs.~(\ref{recast}) and (\ref{marginal}), in the thermodynamical limit
the upper integration extremum diverges for each $E_j$ while, for the marginal probability 
distribution, one has $P_{n} (E_j,E)\simeq {\rm e}^{-\frac{E_j-2}{T}}/T$. 
in the thermodynamical limit, the decoupled energies are distributed according to 
independent Boltzmann distributions, with the parameter $T$ playing the role of a temperature,
in compliance with the equipartition theorem and equivalence of statistical ensembles 
of classical thermodynamics.
This argument shows that the micro-canonical measure fulfills the general canonical principle.
Also, it naturally brings us to introduce a `canonical' measure on the set of pure 
Gaussian states.

\subsection{Canonical measure}

In the `canonical' approach, we will assume for
the energies $\gr{E}$ a probability distribution ${\rm d}p_c(\gr{E})$ reading
\be
{\rm d}p_c(\gr{E}) = \frac{{\rm e}^{-(|\gr{E}|-2n)/T}}{T^n} \,{\rm d}\gr{E} = 
\prod_{j=1}^{n} \left( \frac{{\rm e}^{-(E_j-2)/T}}{T} \,{\rm d}E_j \right) \; , \label{canonic}
\ee
introducing a `temperature' $T$. This distribution maximises the entropy on the knowledge of the 
continuous variables $E_j$'s for given average total energy $E_{av}$, such that $nT=E_{av}$ 
(the latter relation is easily derived by applying Lagrange multipliers).

The `canonical' average $Q_{c}(T)$ over pure Gaussian states at temperature $T$ of 
the quantity $Q(E,\vartheta)$ determined by the second moments alone 
will thus be defined as 
\be
Q_{c}(T) = \int {\rm d}\mu_{H}(\vartheta) \int \frac{{\rm e}^{-(|E|-2n)/T}}{T^n} 
\,{\rm d}E\, Q(E,\vartheta) \; ,  \label{cano}
\ee
where the integration over the energies is understood to be carried out over the whole 
allowed domain ($E_j\ge 2$ $\forall$ $j$).

As already elucidated, the micro-canonical and canonical approaches 
coincide in the thermodynamical limit, as one should expect in analogy with the indistinguishability 
of the classical statistical ensembles in the thermodynamical limit.


\section{Concentration of measure}\label{sec:conce}

In the present section,
we will study the statistical properties of the entanglement of pure Gaussian states
in the thermodynamical limit 
under the measures introduced in the previous section. 
More specifically, we shall focus on the behaviour of the entanglement of a subsystem 
of $m$ modes (as quantified by the von Neumann entropy of the reduction describing such a subsystem), 
keeping $m$ fixed and letting the total number of modes 
$n\rightarrow\infty$. As we have seen, the two measures coincide in this limit 
(when, in the micro-canonical treatment, the energy -- notably the other extensive quantity in play -- 
diverges as well). It will thus suffice to consider the canonical measure, 
and the micro-canonical averages will be retrieved upon identifying $T\equiv(E-2n)/n$.
In this section, the shorthand notation $\overline{x}$ will stand for the average of the quantity $x$ 
with respect to the canonical state-space measure.
Under the previous assumptions, we will determine the average asymptotic 
entanglement and rigorously prove that the variance of the entanglement tends to zero in the thermodynamical 
limit. The latter property, which had been previewed in Ref.~\cite{nosotros}, 
will be referred to as ``concentration of measure''. 

Let us first recall that the von Neumann entropy $S$ of the $m$-mode reduction is determined 
by the local symplectic eigenvalues $\{\nu_j, \,{\rm for}\, 1\le j\le m\}$ of the reduced 
$m$-mode CM $\gr{\gamma}$ according to: 
\be
S = \sum_{j=1}^{m}{h(\nu_j)} \; , \label{vneu}
\ee
where the entropic function $h(x)$ is defined by \eq{entrofunc}, with the additional proviso 
$h(1)\equiv 0$ (which renders the function continuous). Recall also that the uncertainty principle
reads $\nu_j\ge1$ in terms of the symplectic eigenvalues.

In turn, the symplectic eigenvalues are continuously determined by $m$ symplectic invariants 
$\{\Delta^{m}_{d}\}$, given by the even order coefficients of the characteristic polynomial 
of the matrix $\Omega\sig$ \cite{serafozzi06}.
Let us denote by $g$ the continuous function
connecting the invariants to the von Neumann entropy:
\be
S = g(\Delta^{m}_{1},\ldots,\Delta^{m}_{m}) \; . \label{entrolink}
\ee 
Note also that any symplectic invariant $\Delta_{d}^{m}$ 
is a homogeneous polynomial of order $2d$ in the entries of $\gr{\gamma}$. 

In order to prove the concentration of measure for the entanglement, 
we will show that the distribution induced by the canonical measure 
on the space of the symplectic invariants tends, in the thermodynamical limit, 
to a $\delta$-function centred on their averages (which will also be determined). 
Through \eq{entrolink}, this will 
allow us to infer concentration of measure for the von Neumann entropy 
(and to determine its asymptotic average as well).
Notice that this will also imply concentration of measure for any other entropic measure 
as, for Gaussian states, all such quantities are univocally determined 
by the symplectic invariants \cite{serafozzi06,extremal}. 

Let us consider the CM $\sig$ of the whole system.
Considering the expression (\ref{purecm}) for a pure covariance matrix and 
parametrising the transformation $O\in K(n)$ in terms of the matrices $X$ and $Y$ 
such that $(X+iY)\equiv U\in U(n)$, according to the isomorphism of \eq{isu}, one has 
for the entries of $\sig$:
\bea
\sigma_{jk} &=& \sum_{l} \left(X_{jl}X_{kl}z_l^2 + Y_{jl}Y_{kl}z_l^{-2}\right)\label{expli1} \qquad{\rm for}\; 1\le j,k\le n \, , \\
&&\nonumber\\
\sigma_{jk} &=& \sum_{l} \left(Y_{jl}Y_{kl}z_l^2 + X_{jl}X_{kl}z_l^{-2}\right)\label{expli2} \qquad{\rm for}\; n+1\le j,k\le 2n \, ,\\
&&\nonumber\\
\sigma_{jk} &=& \sum_{l} \left(-X_{jl}Y_{kl}z_l^2 + Y_{jl}X_{kl}z_l^{-2}\right)\label{expli3}
\qquad{\rm for}\; 1\le j,(k-n)\le n \, .\\
&&\nonumber
\eea
As previously remarked, each of the symplectic invariants is a homogeneous polynomial in such entries. 
Let us now consider, for fixed squeezings $z_j$'s, the averages of such polynomials with respect to the 
matrices $X$ and $Y$, distributed according to the Haar measure over $U(n)$. 

To work out averages over the Haar measure we shall make use of some basic properties of the integration 
over the unitary group, derived from simple symmetry arguments. 
The reader is deferred to Ref.~\cite{aubertlam} for a general discussion of such strategies. 
Let us also mention that an alternative way for computing Haar integrals with applications 
to linear optical systems and average entanglement has been discussed in Ref.~\cite{designs}.
In particular, since permutation of the indices is clearly a unitary operation, and since the Haar measure 
is both left- and right- invariant with respect to any unitary operation, the integration over the group 
only depends on the number and multiplicity of the different left and right indices present, but not 
on their specific values. Likewise, the measure is invariant under local phase changes ({\em i.e.}~operations 
represented by diagonal matrices with one eigenvalue equal to $e^{i\phi}$ and all the other 
eigenvalues equal to $1$), which implies that all the $X_{jk}$'s can be swapped with 
$Y_{jk}$'s without affecting the values of the integrals. 
Another consequence of the invariance under local phase changes will be exploited shortly. 

Let us consider the average $\overline{\Delta_{d}^{m}}$ of the invariant $\Delta_{d}^{m}$ 
in the limit $n\rightarrow \infty$. 
This invariant is a polynomial of order $2d$ of elements given by Eqs.~(\ref{expli1})-(\ref{expli3}),
so that each term of the polynomial is a product of sums, of the form 
\be
\prod_{s=1}^{2d}\sigma_{j_sk_s} = 
\prod_{s=1}^{2d}
\left(\sum_{l=1}^{n} W_{j_sk_sl}
\right) 
\, , \label{forma}
\ee
where $W_{jkl}$ are polynomials of order two in $X$ and $Y$ depending on the $z_l$'s and 
determined by Eqs.~(\ref{expli1})-(\ref{expli3}).
Suppose we expand the product occurring on the LHS of \eq{forma}, and 
sort the contribution of the resulting addenda 
according to the multiplicity of the different indexes $l$'s, for $1\le l\le n$.
As already remarked, the integral over the Haar measure of the 
monomials in the matrix elements $X_{jl}$ and $Y_{jl}$, depends only on such multiplicities. 
Elementary combinatorial arguments show that the contributions of leading order in $n$ 
are the ones for which all the indexes $l_k$'s are different from each other, which encompasses $n!/(n-2d)!$ 
terms of the resulting sum: all the other terms can be neglected in the thermodynamical limit.

Indeed, only one of the leading terms is non-vanishing. Whenever a term of the sum 
of \eq{expli3} enters in a monomial where all the $l_k$'s are different, the resulting average is of the form 
$
\overline{X_{j1}Y_{k1} q(X,Y)}
$, 
where the index $1$ has been fixed with no loss of generality (due to permutational invariance), 
while the $q(X,Y)$ is any function of the entries of $X$ and $Y$ where $X_{j1}$ and $Y_{k1}$
do not appear. In terms of the complex unitary matrix $U=X+iY$,
one then has 
$$
4i\overline{X_{j1}Y_{k1} q(X,Y)}=\overline{(U_{j1}+U^{*}_{j1})(U_{k1}-U^{*}_{k1})q(X,Y)}=0 \, ,
$$
because the average has to be invariant under arbitrary left and right phase changes on the 
indexes $j$, $k$ and $1$, but all the `phase-invariant' terms $|U_{1j}|^2$ get cancelled out
[this holds also for $j=k$, as in that case the factor depending on $U_{j1}$
reduces to $(U_{j1}^2-U_{j1}^{*2})$]. 
Moreover, if all the $l_k$'s are different, terms coming from the sums of \eq{expli1}
are identical to those coming from the sums of \eq{expli2}, as $X$'s and $Y$'s indexes can be swapped.
Summing up, the leading term as $n\rightarrow\infty$ of the average $\overline{\Delta^{m}_{d}}$, reads 
\be
\overline{\Delta^{m}_{d}} \simeq \, c(d,m,n) \sum_{{\cal S}^{n}_{2d}} \overline{\prod_{l\in {\cal S}^{n}_{2d}}
(z_{l}^2+\frac{1}{z_{l}^2})}
= c(d,m,n) \sum_{{\cal S}^{n}_{2d}} \overline{\prod_{l\in {\cal S}^{n}_{2d}} E_{l}} \; ,
\ee
where the sum runs over all the possible $2d$-subsets ${\cal S}^{n}_{2d}$ 
of the first $m$ natural integers ({\em i.e.~}over all the possible $n!/(n-2d)!\simeq n^{2d}$ combinations of 
$2d$ integers smaller or equal than $n$, with no repetitions, equal to all the possible combinations
of different indexes), 
$c(d,m,n)$ is a factor depending on $d$, $m$ and $n$ which we shall determine shortly,
and 
the symbol $\simeq$ specifies that the equality holds in the thermodynamical limit. 
Note that, being interested in finite subsystems in the thermodynamical limit, 
we can always assume $2d<n$.
The convenience of the parametrisation of the states through the energies $\{E_k\}$ 
becomes fully apparent in the previous expression.
The coefficient $c(d,m,n)$ is easily determined by considering that, setting $E_{k}=2$ $\forall k$, 
one has $\sig=\id$ regardless of the applied orthogonal transformation, for which
the values of the local invariants $\Delta_{d}^{m}$ are fixed and trivially equal to their averages. 
This yields 
\be
\overline{\Delta^{m}_{d}} \simeq {m \choose d} 
\frac{\sum_{{\cal S}^{n}_{2d}} \overline{\prod_{l\in {\cal S}^{n}_{2d}} E_{l}}}
{(2n)^{2d}} \; . \label{avdelta}
\ee
Notice that the factor $n^{-2d}$ in the normalisation is due to the integration over the Haar measure. 
In fact, such a dependence is common to the average of any monomial of order $2d$ 
in the matrix elements (when non null) \cite{aubertlam}. Also, let us emphasise that 
the previous line of thought applies to any polynomial in the entries of $\sig$. 
The only specific information about the invariants $\Delta_{d}^{m}$ entered in imposing 
the normalisation condition, so that all the structure of the invariants could be `extracted' 
from a trivial situation (one has $\Delta_{d}^{m}={m \choose d}$ for the vacuum \cite{serafozzijosab}).

The same arguments apply to the average $\overline{(\Delta_{d}^{m})^2}$ (as a squared 
invariant is just a polynomial of order $4d$ in the entries of $\sig$), leading to
\be
\overline{(\Delta^{m}_{d})^2} \simeq {m \choose d}^2 
\frac{\sum_{{\cal S}^{m}_{4d}} \overline{\prod_{l\in {\cal S}^{m}_{4d}} E_{l}}}
{(2n)^{4d}} \; . \label{vardelta}
\ee

To prove the concentration of measure for the invariants, 
we have to average over the remaining variables $\{E_{j}\}$. 
As we have seen, both the canonical and the micro-canonical distributions are 
i.i.d. in the thermodynamical limit. For the sake of clarity, let us impose 
the two conditions of invariance under permutations and statistical independence one at a time.
The invariance under permutations of the variables $\{E_{k}\}$ allows one to  
write Eqs.~(\ref{avdelta}) and (\ref{vardelta}) as 
\bea
\overline{\Delta^{m}_{d}} &\simeq& {m \choose d} 
\frac{n!\overline{\prod_{k=1}^{2d} E_{k}}}
{(n-2d)!(2n)^{2d}} \simeq {m \choose d} 
\frac{\overline{\prod_{k=1}^{2d} E_{k}}}
{2^{2d}} +O(\frac1n) \, , \nonumber\\
\overline{(\Delta^{m}_{d})^2} 
&\simeq& {m \choose d}^2 
\frac{\overline{\prod_{k=1}^{4d} E_{k}}}
{2^{4d}} +O(\frac1n) \, , \nonumber
\eea
leading to 
\be
\lim_{n\rightarrow\infty} \left(\overline{\Delta_{d}^{m}}^2 - \overline{(\Delta^{m}_{d})^2}\right) = 
\frac{{m \choose d}^2}{16^d}
\left(\left(\overline{\prod_{k=1}^{2d}E_k}\right)^2-
\overline{\prod_{k=1}^{4d}E_k}
\right) \, .
\ee
Notice that the quenching of the residual terms in this development is due to the integration over 
the Haar measure of the compact symplectic group, which always results in terms of the order 
$n^{-2j}$ for polynomials of order $2j$ in $X$ and $Y$. 
Finally, the statistical independence of the energies $E_{j}$'s requires
$\overline{\prod_{j=1}^{k}E_j}=\overline{E}^k$ $\,\forall k$, such that the previous equation becomes
\be
\lim_{n\rightarrow\infty} (\overline{\Delta_{d}^{m}}^2 - \overline{(\Delta^{m}_{d})^2}) = 0 \; . \label{concdelta}
\ee 
Therefore, the variance of any local symplectic invariant $\Delta_{d}^{m}$ vanishes in the thermodynamical limit. 
The previous argument applies to any polynomial of finite order in the entries of $\sig$. 
In particular, in the case $2d=1$, this vanishing of the variance directly applies to the entries of $\sig$, 
thus also implying, {\em a posteriori}, the compliance (already shown in Ref.~\cite{nosotros} in a more specific 
background) of the presented measures with the general canonical principle.

To complete our proof and extend the concentration of measure to the local von Neumann entropy, 
let us consider the measure $\Gamma_n(\Delta_{1}^{m},\ldots,\Delta_{m}^{m})$ induced, 
in the $m$-dimensional real space of the local symplectic invariants $\{\Delta_{d}^{m}\}$, 
by the proposed canonical measure on $n$-mode pure Gaussian states. 
Because of \eq{concdelta}, such a measure completely 
concentrates in the average values $\overline{\Delta_{d}^{m}}$. 
Here, we will express this fact by claiming that $\Gamma_n$ tends to a Dirac delta:
$$
\lim_{n\rightarrow\infty}\Gamma_n(\Delta_{1}^{m},\ldots,\Delta_{m}^{m}) = 
\delta(\Delta_{1}^{m}-\overline{\Delta_{1}^{m}},\ldots,\Delta_{m}^{m}-\overline{\Delta_{1}^{m}}) \, .
$$
[see \ref{mathematicalize} for a rigorous formulation, 
showing the emergence of the following limits from Eq.~(\ref{concdelta})].
For the von Neumann entropy of the $m$-mode subsystem $S_m$, one finds
\be
\fl\lim_{n\rightarrow\infty} \overline{S_m} = 
\lim_{n\rightarrow\infty} \int \Gamma_n(\Delta_{1}^{m},\ldots,\Delta_{m}^{m}) g(\Delta_{1}^{m},\ldots,\Delta_{m}^{m})
\,{\rm d}^{m}\Delta_{d}^{m} 
= g(\overline{\Delta_{1}^{m}},\ldots,\overline{\Delta_{m}^{m}}) \; , \label{avvneu}
\ee
\be
\lim_{n\rightarrow\infty} \overline{S_m^2}-\overline{S_m}^2 = 0 \; . \label{varvneu}
\ee

The asymptotic average $\overline{S_m}$ can be determined as well. 
First, let us determine the canonical averages $\Delta_{d}^{m}$ of the invariants 
according to Eqs.~(\ref{avdelta}) and (\ref{canonic}), finding
\be
\lim_{n\rightarrow\infty}\Delta_{d}^m = {m\choose d} \left(1+\frac{T}{2}\right)^{2d} \; . \label{canavdelta}
\ee
Next, let us recall the expression of the symplectic invariants in terms of the symplectic eigenvalues 
$\{\nu_{j}\}$ \cite{serafozzi06}:
\be
\Delta^m_d = 
\sum_{{\cal S}^{m}_{d}} \prod_{k\in{\cal S}^{m}_{d}} \nu_{k}^{2} \; .\label{inva}
\ee
Straightforward substitution in \eq{inva} shows that the values $\nu_k=(1+T/2)$, $\forall 1\le k\le m$, 
account for the asymptotic values of the invariants given by \eq{canavdelta}. 
Thus, for the symplectic eigenvalues $\{\nu_j\}$, one has
\be
\lim_{n\rightarrow\infty} \overline{\nu_{k}} = 1+\frac{T}{2} \quad \forall \, 1\le k \le m \; .
\ee
In the thermodynamical limit, both the introduced distributions, canonical 
and micro-canonical, comply with a form of equipartition theorem: an average 
`energy' equal to $T/2$ is allotted to each decoupled quadratic degree of freedom.
Finally, one has 
\be
\lim_{n\rightarrow\infty} \overline{S} = m h(1+\frac{T}{2}) \; . \label{asentro}
\ee

\section{Study of the typical entanglement of pure Gaussian states}\label{study}

In the present section, we present a detailed study of the typical bipartite entanglement for Gaussian states 
with a finite number $n$ of total modes. More specifically, we will study the statistical 
properties of the entropies of a reduced subsystem of $m$ modes, when the global states are 
distributed according to the micro-canonical and canonical measures.
As we have already mentioned, the quantity $\det{\sig}$ is a proper quantifier of the {\em mixedness} 
of Gaussian states, and thus of their entanglement as well, when computed for the 
reduced subsystem of a pure state. 
For such an entropic quantity, we were able to derive analytical expressions for the canonical and 
micro-canonical average and standard deviation, which will be presented in subsection \ref{sec:purity}. 
In subsection \ref{sec:bounds} we describe a strategy -- based on a simple linear minimisation -- to derive, 
from such analytical results, exact information about the micro-canonical statistical properties of the actual entropy 
of entanglement (defined as the von Neumann entropy of the subsystem). 
In subsection \ref{sec:numerics} we complement this analysis with numerical results,
also covering $m$-mode subsystems.

\subsection{Purity of single-mode subsystems}\label{sec:purity}

For a single-mode subsystem (corresponding to $m=1$ in the previous section's notation),  
we have worked out analytically
the average and variance of the inverse squared purity $\mu^{-2} = 1/\tr[\varrho^2]^2$,
which coincides, for Gaussian states, with the determinant of the reduced covariance matrix $\gr{\gamma}$.
Let us mention that, in this particular instance, such a measure is a perfectly legitimate entanglement quantifier,
as it induces, on the set of single-mode states, the same hierarchy as the von Neumann entropy
(in fact, both quantities are determined by the single symplectic eigenvalue, see Section \ref{techintro}) 
\cite{serafozzijpb}.

Making use of the methods described in Ref.~\cite{aubertlam}, we determined 
the first and second statistical moments over the Haar measure 
(respectively denoted by ${\mu^{-2}_H}$ and ${\mu^{-4}_H}$) 
of the determinant $\det{\gr{\gamma}}=\mu^{-2}_H$, to find
\be
{\mu^{-2}_H} = \sum_{j\neq k}\frac{{E_j E_k}}{4(n+1)n} + \frac{2}{n+1} \, ,
\ee
\bea
\fl{\mu^{-4}_H} &=&  \frac{1}{16}\frac{(n-1)!}{(n+3)!} \nonumber
\left[ \sum_{j\neq k\neq l \neq m}{E_jE_kE_lE_m}
+ 8\sum_{j\neq k\neq l}{E_j^2E_kE_l} + 12\sum_{j\neq k}{E_j^2E^2_k} \right. \\
\fl&&\left. +\left(96+16(n-2)\right) \sum_{j\neq k}{E_jE_k} -32(n-1)\sum_{j}{E_j^2} 
+ 128n(n-1)+384n \right] ,
\eea
where the averages over the variables $\{E_{j}\}$ are still to be worked out, 
according to the chosen distribution.

The canonical and micro-canonical averages are then straightforward to compute. 
In the canonical instance, the average and second moment $\mu^{-2}_{c}$ and $\mu^{-4}_{c}$ read:
\be
\mu^{-2}_{c} = \frac14 \frac{n-1}{n+1}(T^2+4T) + 1  \, ,
\ee
\bea
\fl{\mu^{-4}_{c}} = \frac{1}{16}\frac{n! (n-1)}{(n+3)!}
\Bigg[(n^2+11n+22) T^4 +8(n^2+8n+6) T^3 +8(3n^2+15n+10) T^2 \nonumber\\
 +32 (n+3)(n+2) T \Bigg] +1 .
\eea

Whereas, in the micro-canonical case, defining $\tilde{E} \equiv (E-2n)$, one gets
\be
\mu^{-2}_{mc} = 
\frac{(n-1)}{4(n+2)(n+1)^2}\Big(\tilde{E}^2+4(n+2)\tilde{E}\Big) + 1 \, ,
\ee
\bea
\fl\mu^{-4}_{mc} &=& 
\frac{(n!)^2(n-1)}{16(n+4)!(n+3)!}\left( (n^2+11n+22)\tilde{E}^4 +8(n+6)(n+4)(n+1)\tilde{E}^3 
\right. \nonumber \\
\fl&&\left. + 8(n+4)(n+3)(3n^2+15n+10)\tilde{E}^2 +32(n+4)(n+3)^2(n+2)^2\tilde{E} \right) + 1 .
\eea
Notice that the maximal $\mu^{-2}_M$ for given energy $E=\tilde{E}+2n$ is
(see \ref{maximal})
\be
\mu^{-2}_M = \frac{(\tilde{E}+4)^2}{16} \; . \label{minpur}
\ee

Whilst restricted to a particular quantifier and to a single mode, these analytical results 
display the most relevant statistical features of the entanglement of pure Gaussian states. 
The micro-canonical mean $\mu_{mc}^{-2}$ is monotonically increasing with $E$ for fixed $n$ and, 
for $n>2$, monotonically decreasing with $n$ for given $E$ (a ``finite size'' effect shows up for 
$\tilde{E}\le10$, where $\mu_{mc}^{-2}$ increases in going from $2$ to $3$ modes).
This can be promptly explained as more available energy generally allows for higher entanglement, 
while the presence of more modes ``drains'' energy away to establish correlations which do not involve the 
particular chosen mode.
On the other hand, the canonical average entanglement is monotonically increasing 
in both the temperature and the number of modes.
This behaviour is encountered also for the micro-canonical entanglement 
with given maximal total energy {\em per mode}, which is, even for small $n$, 
closely akin to the canonical ensemble (upon replacing $\tilde{E}/n$ with $T$).
The increase of the average canonical entanglement with increasing number of modes but 
fixed temperature is a non trivial, purely `geometric' effect, 
due to the average over the Haar measure ${\rm d}\mu_{H}(\vartheta)$
of the compact variables $\vartheta$. 
An analogous increase is in fact observed assuming a given, fixed value for the variables $z_{j}$'s 
and averaging only over the compact variables $\vartheta$: as the number of total modes increases, 
a given mode has more possibilities of getting entangled, even keeping a fixed mean energy per mode.

As for the standard deviations, which are straightforward to derive from the expressions above, 
they are generally increasing with total energy and temperature for fixed total number of modes 
(as more energy allows for a broader range of entanglement). 
Significantly, these partial analytical results clearly show the arising of the concentration of measure 
around a thermal average. Both for the canonical case and for the micro-canonical one with 
$\tilde{E}=nT$ the standard deviation decreases with increasing number of modes $n$, 
falling to zero asymptotically
(after transient ``finite size'' effects, for very small $n$). 
Moreover, in the micro-canonical instance, the thermal average 
of concentration is generally very distant, even for relatively small $n$, from the allowed maximum 
of \eq{minpur} (which clearly diverges in the thermodynamical limit): {\em e.g.}, 
for $\tilde{E}=10n$ one has that the average $\mu_{mc}^{-2}$ is, respectively $16.5$  
and $257.1$ standard deviations away from the maximal value $\mu^{-2}_{M}$ for
$n=5$ and $n=20$. Such a distance increases monotonically with the total number of modes.
This peaked concentration for finite $n$ will be exploited in the next subsection 
to obtain strict bounds on the average von Neumann entropy of entanglement of single-mode 
subsystems.


\subsection{Estimating the micro-canonical mean entropy of entanglement}
\label{sec:bounds}

In the previous section, we have focused on the inverse square purity $\mu^{-2}$, because it
can be readily described in terms of the covariance matrix's entries, 
thus allowing one to determine analytical expressions for the entanglement's
statistics.
Let us stress once more that, for pure states,
such a quantity is a legitimate entanglement measure, as it is a monotone 
strictly related to the so-called linear entropy, given by $1-\mu$.
Moreover, for single-mode Gaussian states, the linear entropy is, 
as we have already remarked, monotonically determined by the von Neumann entropy alone.
However,
the proper ``entropy of entanglement'' (given by the von Neumann entropy of the reduced density matrix)
is endowed with a clear
operational meaning for pure states (as it corresponds to the rate of singlets distillable 
from the state in the asymptotic limit of infinite copies \cite{bennett}). 
It is thus highly desirable to obtain qualitative and quantitative information 
about the statistical properties of such a quantity under our measures. 
Clearly, the findings of Section \ref{sec:conce} already
provide us with such results in the thermodynamical limit, where the von Neumann entropy concentrates, 
with vanishing variance, around the value set out in \eq{asentro}. 

Here, we shall address the entropy of entanglement of a single-mode subsystem in a system 
with a finite total number of modes. Exploiting the fact that the
von Neumann entropy $S$ is continuously determined by $\mu^{-2}$ as \cite{serafozzijpb}
\be\label{eqn:entropyByPurity}
        S(\mu)
        = h(\mu^{-1}) \nonumber
\ee
(under the additional prescription $S(1)\equiv0$), we will show how upper and lower bounds 
on the micro-canonical average $S_{mc}$ can be derived. 
The reason why we focus on the micro-canonical measure is that, imposing a bound on the energy, 
it involves a maximal value for the entanglement (as discussed in \ref{maximal}), 
which is another key ingredient we are going to use. 
Extending the method we are presenting to the canonical measure is possible, 
at the price of introducing an approximation to neglect the `tail' of the canonical distribution 
in the space of the entanglement (certainly feasible with reasonable resources for small enough temperatures).
As we will see, such bounds, which we know to become increasingly close with increasing $n$ 
because of concentration, can be quite tight even for relatively small $n$, thus providing precise information 
about the average entropy of entanglement. 

For the sake of readability, we will from now on set $a:=\mu^{-2}$.
For a given number $n$ of modes and total energy $E$ in the system, 
the maximum value $a_{\max}$ for the inverse squared purity of any fixed mode 
is given by \eq{minpur}, while the minimum value is always (for $n>2$) given by $a_{\min}=1$, 
corresponding to the case where the mode is decoupled from the remainder of the system.
Our measure on Gaussian states will induce a probability
distribution $\nu$ on the interval $[a_{\min}, a_{\max}]$. While we do
not know $\nu$ directly, we are aware of two of its properties, namely
the averages $a_c$ and $\overline{a_c^2}$ with respect to it, which we computed in the previous subsection.
Our approach is going to be as follows: to obtain an upper bound for
$S_{mc}$ we maximise $S_{mc}$ over all
probability distributions $\nu'$ which produce the right averages for
$a$ and $a^2$. There is a technical obstacle to the pursuit of this
programme: the set of all probability distributions on the
interval is infinite-dimensional. To circumvent this problem, we will
partition $[a_{\min},a_{\max}]$ into $M$ equally sized sub-intervals
($M\in\mathbbm{N}$ being an arbitrary parameter) and consider
``discretised'' probability distributions, which are constant over these subintervals.
This is going to be done in such a manner that the resulting bounds
are valid for any finite $M$, and not only in the limit of
$M\to\infty$. Also, it will turn out that all optimisations can be
cast into the form of linear programs. This excludes the occurrence of local minima
and implies that the obtained bounds will hold rigorously. 

\begin{figure}[t!]
\begin{center}
\includegraphics[scale=1]{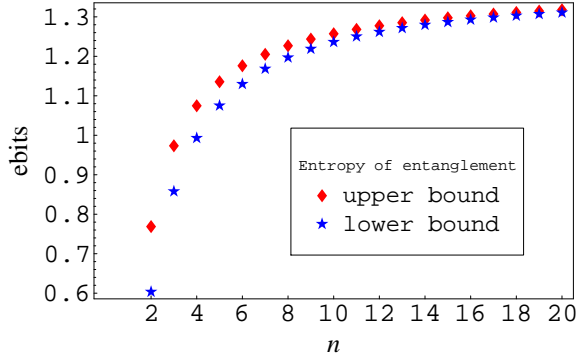}
\caption{
        \label{fig:bounds}
        Upper and lower bounds to the micro-canonical average von Neumann entropy of
        entanglement of one mode as a function of the number of total modes $n$. 
        The maximal energy was assumed to be $4n$ (in units of $\hbar\omega/4$).
        To compute the bounds, the method described in subsection
        \ref{sec:bounds} has been employed. The parameter $M$ was chosen to be $10000$.}
\end{center}
\end{figure}

More explicitly, fix an $M\in\mathbbm{N}$. The $k$th sub-interval will
be $[l(k),l(k+1)]$, where $l(k)\equiv a_{\min}+k \frac{a_{\max}-a_{\min}}{M}$
is its leftmost point. Let $\nu_k:=\nu([l(k),l(k+1)]$ be the measure
of the $k$th sub-interval. 
Clearly, one has
\begin{eqnarray}
        a_{mc} 
        &=&
        \int_{a_{\min}}^{a_{\max}} a\, {\rm d}\nu(a)
        \leq \sum_{k=0}^{M-1} l(k+1) \nu_k,  \quad
        a_{mc} \geq \sum_{k=0}^{M-1} l(k) \nu_k \; , \label{eqn:const1} \\
        a^2_{mc} 
        &\leq&
        \sum_{k=0}^{M-1} l(k+1)^2 \nu_k,  \quad
        a^2_{mc} \geq \sum_{k=0}^{M-1} l(k)^2 \nu_k \, .\label{eqn:const2}
\end{eqnarray}
Furthermore, because $S(a)$ is monotonously increasing in $a$, we have
\be
        S_{mc} 
        =
        \int_{a_{\min}}^{a_{\max}} S(a) \,{\rm d}\nu(a) 
        \leq 
        \sum_{k=0}^{M-1} S(l(k+1)) \nu_k \, . \label{eqn:objective1}
\ee
Now let $P\subset \mathbbm{R}^M$ be the set of probability
distributions which are constant on the $M$ sub-intervals and
compatible with Eqs. (\ref{eqn:const1}), (\ref{eqn:const2}). 
Certainly, the discrete distribution given by the $\nu_k$ is an element
of $P$ and we can thus deduce from Eq. (\ref{eqn:objective1}) that
\begin{eqnarray}\label{eqn:lin1}
        S_{mc} 
        &\leq&
        \sup_{\nu' \in P} 
        \sum_{k=0}^{M-1} S(l(k+1)) \nu'_k \, .
\end{eqnarray}
Hence, the following linear program with variables
$\nu'\in\mathbbm{R}^M$ yields an upper bound for
$S_{mc}$:
\begin{eqnarray*}
        {\rm maximise} && 
        \sum_k S(l(k+1)) \nu_k \\
        {\rm subject}\;{\rm to} \quad
        && a_{mc} \leq \sum_k l(k+1) \nu'_k\, , \quad
         a_{mc} \geq \sum_k l(k) \nu'_k\, ,  \\
        && a_{mc}^2 \leq \sum_k l(k+1)^2 \nu'_k\, , \quad
         a^{2}_{mc} \geq \sum_k l(k)^2 \nu'_k\, , \\ 
        && \sum_k \nu'_k = 1.
\end{eqnarray*}
A lower bound is found by a completely analogous minimisation program.
        
Figure \ref{fig:bounds} depicts the results of the programme laid out
above for a specific choice of energy per mode. Quite remarkably, 
already for a small number of modes, one can give a fairly
precise value for the average micro-canonical entropy of entanglement, 
thus complementing the
asymptotic statement of Section \ref{sec:conce}.
Let us also mention that the restriction imposed by the knowledge of the second moments 
is crucial in rendering the bounds so tight (as opposed to the use of the averages alone).

\begin{figure}[t!]
\centerline{
\includegraphics[scale=.58]{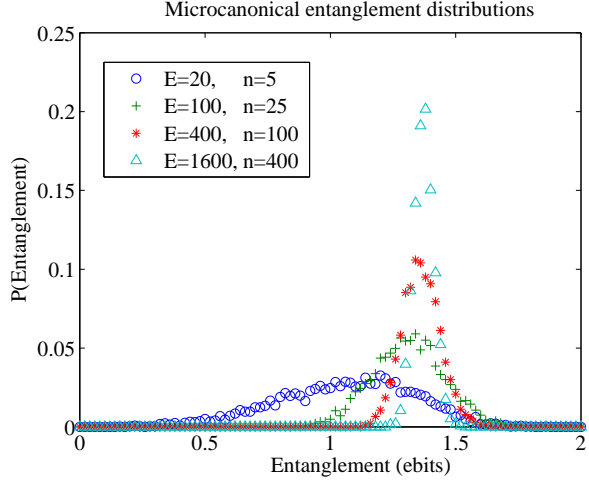}}
\vspace{.0cm}
\caption{Histograms of microcanonical entanglement distributions. $m=1$, and 5000 samples were taken
for each distribution. The concentration of measure with increasing $n$ is apparent.}
\label{fig:Gauhistograms}
\end{figure}

\subsection{Numerical results on the entropy of entanglement}\label{sec:numerics}

Unfortunately, the analytical and exact methods illustrated above cannot be easily 
extended to more complicated situations. 
However, our measures are well suited to be numerically investigated by direct sampling
(the Haar measure-distributed symplectic orthogonal can be reproduced by generating unitary matrices from 
the Gaussian Unitary Ensemble \cite{metha} and then by translating them according to \eq{isu}).\footnote{The 
MATLAB code we made use of is available at {\sf \scriptsize www.imperial.ac.uk/quantuminformation}.}
This allows for the investigation of the statistical properties of the actual entropy of entanglement 
for varying values of $m$, $n$ and of the parameters of the measures $E$ or $T$, 
according to the setting in question.

In Figure \ref{fig:Gauhistograms}
a sequence of numerically generated microcanonical probability distributions for the entanglement of 
a single mode are plotted, unambiguously showing the concentration of measure for 
the von Neumann entropy at small $n$.
Notice that, in the microcanonical case, 
even for small $n$ -- well before the onset of thermodynamical concentration of measure around 
the finite thermal average -- the entanglement of pure Gaussian states distributes 
around values generally distant from the finite allowed maximum: {\em
e.g.}, for $m=1$ and $E= 10n$, the difference between the maximum
and the average $\overline{S}$ is, respectively, 
$4.0$ and $13.6$ standard deviations for $n=5$ and $n=20$.

\begin{figure}[t!]
\centerline{
\includegraphics[scale=0.5]{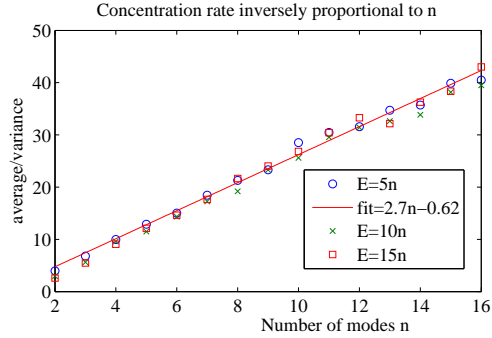}
}
\vspace{.0cm}
\caption{Numerical evidence that the ratio variance/average, for a microcanonical measure 
on $n$-mode states with maximal energy $E$, is proportional to $1/n$ and, 
surprisingly, that the proportionality factor is independent of $E/n$. 
A single-mode subsystem was considered ($m=1$);
1500 states were sampled for each data point.}
\label{fig:inverseN}
\end{figure}

Furthermore, for the micro-canonical case,
our investigation through sampling 
indicates that the ratio between variance and average is
inversely proportional to $n$ for finite $n$ and, rather surprisingly, 
that the proportionality factor is independent of $E/n$ (see Fig.~\ref{fig:inverseN}). 

\begin{figure}[t!]
\begin{center}
\subfigure[] {
\includegraphics[scale=0.33]{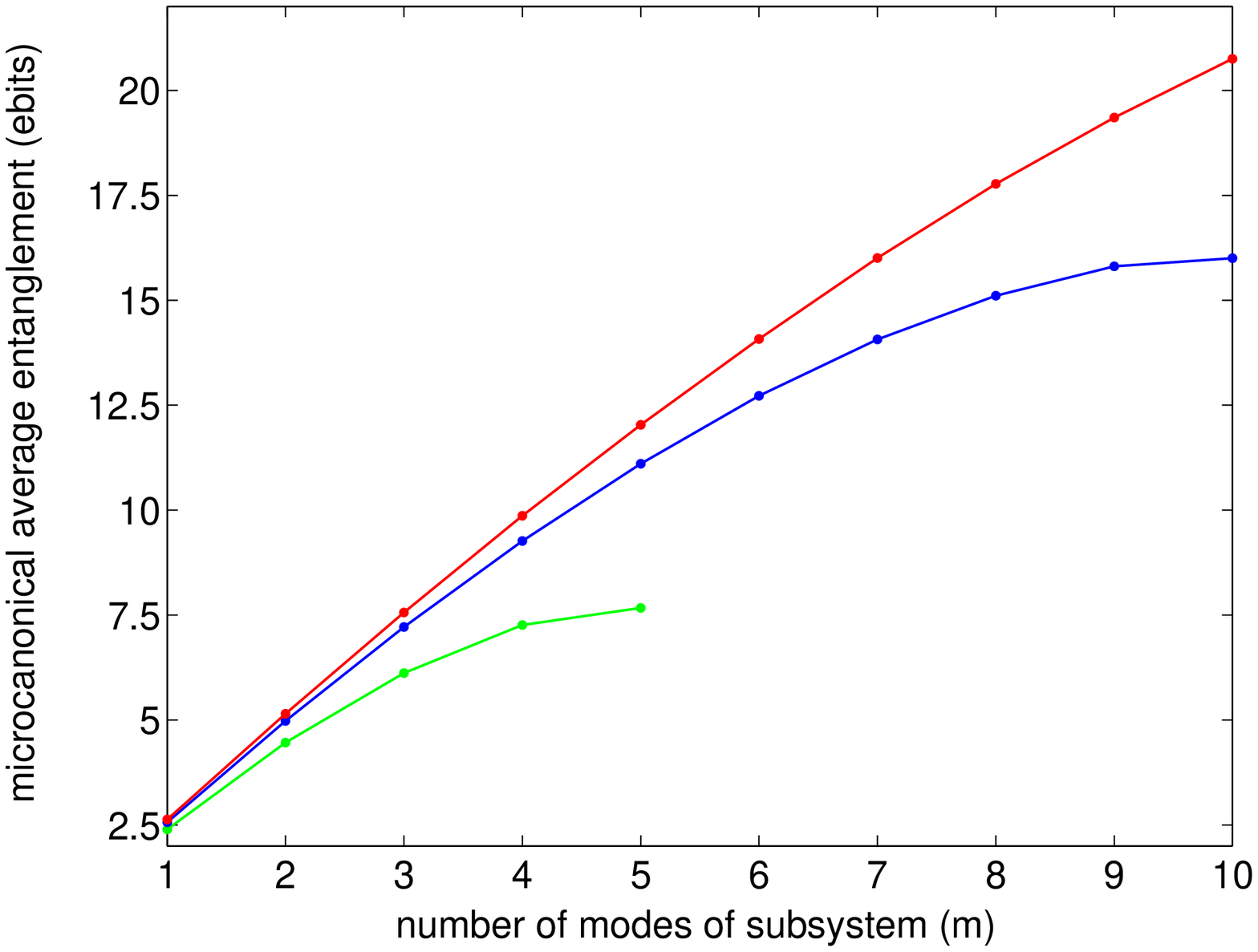}\label{MdepAv}}\hspace*{.2cm}
\subfigure[] {
\includegraphics[scale=0.33]{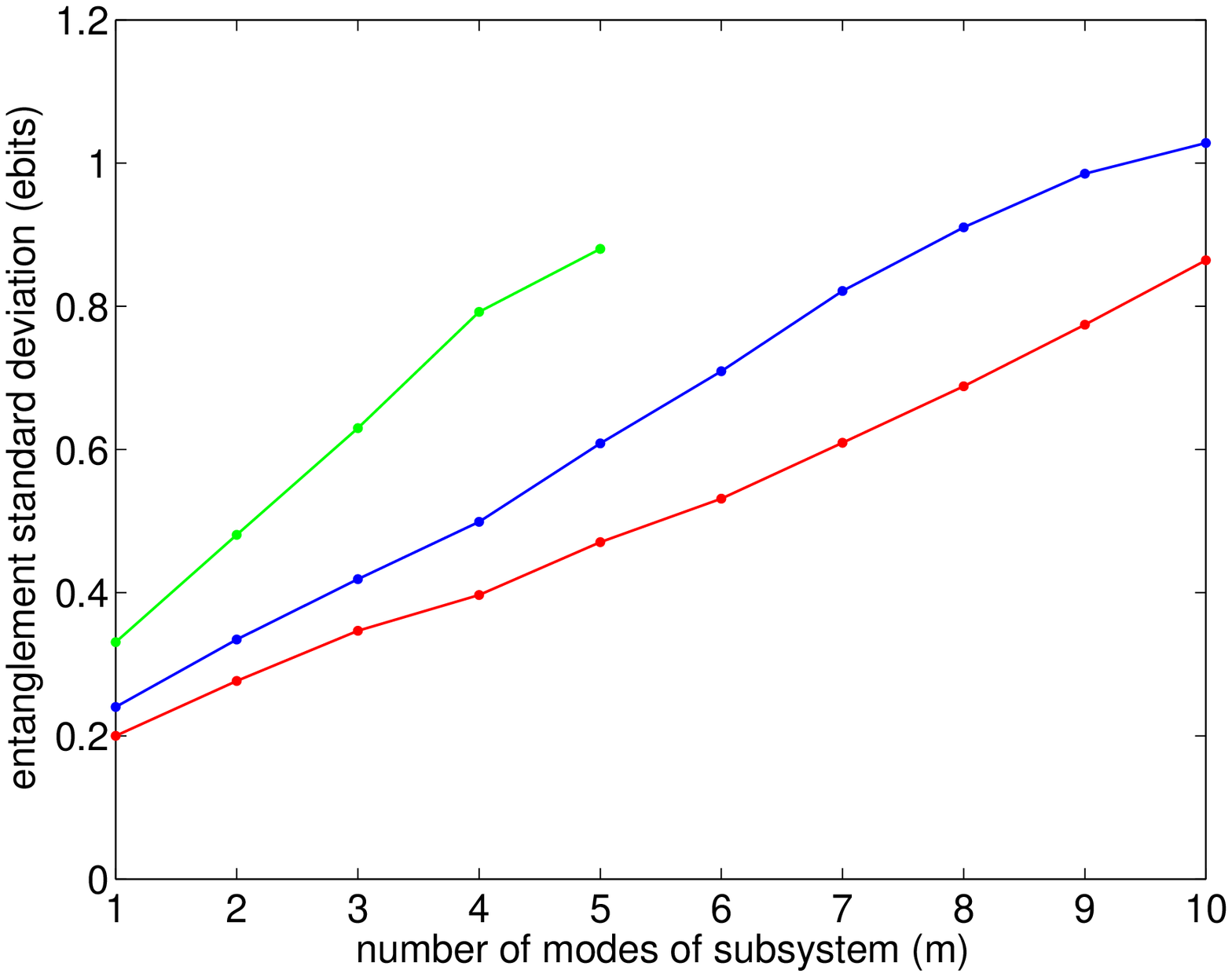}\label{MdepStdev}}
\caption{Microcanonical average entropy of entanglement (a) and standard deviation of the 
entropy of entanglement's distribution (b) for $E=10n$, as a function of the number of modes $m$ of the reduced 
subsystem. On the left (a), the upper curve (red) refers to $n=30$, the middle curve (blue) to $n=20$ and 
the lower curve (green) to $n=10$.
On the right (b), the upper curve (green) refers to $n=10$, the middle curve (blue) to $n=20$ and 
the lower curve (red) to $n=30$.
Only the range $1\le m \le 5$ was considered for $n=10$.
Data were obtained from samples of $5000$ states.
\label{fig:Mdep}}
\end{center}
\end{figure}

When more than one mode is addressed ({\em i.e.} when $m>1$), 
numerics show that the average 
canonical entanglement is roughly proportional to the number of local modes $m$. 
The same linear approximation carries over to the micro-canonical case
for $E\gg n$ and $m\ll n$, upon substituting $E/n$ for $T$.
Clearly, such an approximate proportionality, rigorously true in the 
thermodynamical limit [see \eq{asentro}], is better satisfied for increasing number of total modes $n$, 
but provides good estimates (up to some percent) already for $n\approx 30$ and small enough $m$, 
as is shown in Fig.~\ref{MdepAv}, where a clear subadditive behaviour 
for $m\lesssim n$ is also apparent. 
Also, let us point out that the concentration of measure for the entanglement distribution 
clearly shows up at finite $n$ for $m>1$ as well. For instance, for a microcanonical 
distribution with $m=5$, $n=20$ and $E=200$, one finds that the average is 
$11.4$ standard deviations away from the allowed maximum. Also, Fig.~\ref{MdepStdev} shows 
how the standard deviation increases approximately proportionally to $m$ and decreases 
for increasing $n$ for $m\neq 1$ as well.

\subsection{Concentration of measure for infinite subsystems}
So far, we have considered the concentration of measure occurring for a finite subsystem 
with a fixed number of modes $m$. 
Let us now briefly turn to the case of a fixed {\em `ratio'} of subsystems, where 
$m/n\simeq \alpha>0$ in the thermodynamical limit. 
Clearly, in such a case, the subsystem as well 
is comprised of infinitely many modes

The analytical reasoning of Section \ref{sec:conce} can be readily adapted to this case and
shows, not surprisingly, that the average $\overline{S}$ of the von Neumann entropy diverges 
in the thermodynamical limit.
In this case though, analogous arguments imply that 
the variance $\overline{S^2}-\overline{S}^2$ of the von Neumann 
entropy diverges as well. 
Still, numerics strongly suggest the occurrence of a weaker form of concentration, namely 
\be
(\overline{S^2}-\overline{S}^2)/\overline{S}^2\simeq 0 \, . \label{weakconc}
\ee
This behaviour, 
unambiguously supported by the numerical analysis 
in both the micro-canonical and canonical instances, extends to the 
general case of large $m$ the evidence depicted in Fig.~\ref{fig:inverseN}
(according to which the ratio between variance and average is inversely
proportional to $n$). 

\section{Conclusions and outlook}\label{outlook}

The approach we have introduced here tames the divergence due to the infinite dimension of the Hilbert 
spaces of continuous variable systems and allows one to introduce a well defined notion 
of ``typical'' continuous variable entanglement. 
Our exhaustive analytical and numerical study shows that such a typical entanglement concentrates 
sharply around a thermal average, even for a relatively small number of modes. 
In the micro-canonical case, where the upper bound to the energy implies the existence of an absolute upper bound 
for the entanglement, the 
average entanglement turns out to be different and well separated 
from the maximal value, by many standard deviations, 
even for very small number of modes. 
Under such `heavily thermodynamical' prescriptions -- as are the ones we have adopted to construct 
the canonical and micro-canonical measures -- 
equipartition prevents the entanglement 
from reaching the allowed maximum, evenly spreading the correlations between all the modes.

Our measures, being compliant with the general canonical principle, should be suitable to the description 
of dynamical situations \cite{gemmer,verhulst}, also involving randomised interactions \cite{oliveira}. 
A systematic study of such processes 
-- notably restricting to two-mode interactions, in the spirit of Ref.~\cite{oliveira} -- 
where the two measures defined here would arise as stationary distributions, 
is the next direction to pursue in the present line of research.

This line of enquiry may also contribute to our understanding of why objects typically appear classical despite 
being governed, as is normally assumed, by quantum mechanics. In particular, 
this and other related works, like Refs.~\cite{hayden,popescu05,gemmer},
imply that, always in a particular sense, most global pure quantum states have the property that a local state 
of a comparatively small party, obtained by tracing over the larger party, 
will be highly mixed and accordingly have little or no quantum correlations. 
This is at least one necessary ingredient for the system to appear classical. 
The work here indicates that these arguments can be applied in the continuous variable setting too. 
It would be interesting to combine, compare and further develop existing approaches 
in order to explain the emergence of apparent classicality in realistic models of physical systems. 
As a further investigation, one could consider the extension of
analogous micro-canonical and canonical measures on general finite-dimensional 
states \cite{verhulst}. 
One could then distinguish, to a greater detail, 
the features induced by the chosen thermodynamical setting from the 
ones proper to the Gaussian continuous variable scenario.

Finally, let us mention that the presented framework may be suitably extended to 
more general, {\em mixed} Gaussian states. In fact, the Williamson decomposition of a generic 
covariance matrix [\eq{williamson}], 
together with the Euler decomposition given by \eq{euler1},
suggest that, to encompass mixed states as well, one has to add only another set of compact variables 
$\vartheta$, plus the symplectic eigenvalues of the global mixed states $\{\nu_j\}$. 
The presented approach thus also paves the way for 
the definition of more general measures on the whole set of Gaussian states.

\appendix

\section{A Baker-Campbell-Hausdorff formula for quadratic polynomials in the canonical operators}\label{bch}

For the sake of completeness and self-consistency, we present here a proof of \eq{pures2}. 
To this end, we will follow a strategy customarily adopted in the derivation of Baker-Campbell-Hausdorff--like 
relations (see, {\em e.g.}, Ref.~\cite{barnradmore}).
Let us thus define the operator $\hat{G}_{A,b}(\vartheta) \equiv 
\,{\rm e}^{i\vartheta(\hat{R}^{\sf T} A \hat{R}+\hat{R}^{\sf T}b)}$, 
where $\vartheta$ is a real variable.
Note that the CCR straightforwardly imply 
\be
[\hat{R}^{\sf T} A \hat{R} , \hat{R}^{\sf T}b] = 4iA\Omega b \; . \label{acr}
\ee
We start by assuming the following ansatz, which we will prove shortly,
\be
\hat{G}_{A,b}(\vartheta) \equiv
\,{\rm e}^{i\vartheta(\hat{R}^{\sf T} A \hat{R}+\hat{R}^{\sf T}b)} =
\,{\rm e}^{i f(\vartheta)(\hat{R}^{\sf T} A \hat{R})}\,{\rm e}^{\hat{R}^{\sf T}(M(\vartheta)b)} \, ,
\ee
where $f(\vartheta)$ is a scalar real function, whereas $M(\vartheta)$ is a $2n\times2n$ matrix depending 
continuously on $\vartheta$.
Differentiating both sides with respect to $\vartheta$, we find (differentiation is denoted by $'$):
\bea
\fl-i\frac{{\rm d}\,\hat{G}_{A,b}}{{\rm d}\,\vartheta}(\vartheta) &=&
\left(\hat{R}^{\sf T} A \hat{R}+\hat{R}^{\sf T}b\right)\,\hat{G}_{A,b}(\vartheta) \label{1row} \\
&=&{\Big(f'(\vartheta)\hat{R}^{\sf T} A \hat{R}} + 
\hat{R}^{\sf T}\,{\rm e}^{i f(\vartheta)(\hat{R}^{\sf T}A\hat{R})}
M'(\vartheta)b\,{\rm e}^{-i f(\vartheta)(\hat{R}^{\sf T}A\hat{R})}\Big)
\,\hat{G}_{A,b} \nonumber\\
&=&{\Big(f'(\vartheta)\hat{R}^{\sf T} A \hat{R}} + \hat{R}^{\sf T}
\,{\rm e}^{-4f(\vartheta)A\Omega}
M'(\vartheta)b\Big)
\,\hat{G}_{A,b} \, , \label{3row}
\eea
where 
we have made use of \eq{acr} in the last step.
Equating (\ref{1row}) and (\ref{3row}) yields the following systems of differential equations:
\bea
f'(\vartheta) &=& 1 \; , \\
M'(\vartheta) &=& {\rm e}^{4f(\vartheta)A\Omega} \; ,
\eea 
with initial conditions $f(0)=0$ and $M(0)=0$. This system always admits an analytical solution, 
given in general by $f(\vartheta)=\vartheta$ and 
$M(\vartheta)=\int_{0}^{\vartheta} {\rm e}^{4\vartheta'A\Omega} {\rm d}\vartheta'$, 
whose value in $\vartheta=1$ gives the matrix $M=M(1)$, thus proving the validity of
\eq{pures2}. 
If $A$ is invertible, $M$ is simply given by
\be
M = \frac14 \Omega A^{-1} \left(\id_{2n} 
- \,{\rm e}^{4A\Omega}\right) \; .
\ee

\section{Asymptotic concentration of measure}\label{mathematicalize}

In this appendix, we show how the concentration of the measure $\Gamma_{n}$ in the space of the 
symplectic invariants follows from \eq{concdelta}. In doing so, 
we will provide the reader with a formal derivation of Eqs.(\ref{avvneu}) and (\ref{varvneu}).

Let us consider the $m$-dimensional real space $\Delta^{m}$ of the vectors 
of symplectic invariants $\Delta\equiv(\Delta_1^{m},\ldots,\Delta_{m}^{m})$, 
endowed with the usual Euclidean norm $\|\cdot\|$. 
Let $D_{\varepsilon}$ be a 
spherical ball of radius $\varepsilon$  
centered in $\overline{\Delta}\equiv(\overline{\Delta_{1}^{m}},\ldots,\overline{\Delta_{m}^{m}})$:
$D_{\varepsilon}=\{\Delta \,:\, \|\Delta-\overline{\Delta}\|\le \varepsilon\}$.
Let $R_{\varepsilon}$ be the complement of $D_{\varepsilon}$: 
$R_{\varepsilon}=\{\Delta \,:\,\|\Delta-\overline{\Delta}\|> \varepsilon\}$. 
Recall that $\Gamma_{n}$ stands for the (normalised) measure induced on the space 
$\Delta^{m}$ by the canonical measure of $n$-mode pure Gaussian states. 
Also, $\Gamma_{n}({\rm d}\Delta)$ will stand for the infinitesimal element of such 
a measure.
Let us remark that \eq{concdelta}, holding $\forall\,d$, implies
\be
\lim_{n\rightarrow\infty} (\overline{\|\Delta\|^2} - \|\overline{\Delta}\|^2) = 
\overline{\|\Delta-\overline{\Delta}\|^2} = 0 \; . \label{concdeltoso}
\ee

Our first aim is deriving a rigorous formulation of ``concentration of measure'', {\em i.e.}: 
\be
\forall \varepsilon>0 \;{\rm and}\; \forall \xi>0\, , \;\;\exists\;\tilde{n} \,|\, 
\,\forall\, n>\tilde{n} \,:\,\Gamma_{n}(R_{\varepsilon})<\xi \, .\label{conce}
\ee 
To this aim suppose, {\em ad absurdum}, that the latter statement did not hold. 
Then, $\forall\,n$, $\exists\, \varepsilon,\xi>0$ and $n_{0}>n$ such that one has
$\Gamma_{n_0} (R_{\varepsilon})>\xi$. But this would imply that, $\forall\, n$, $\exists\,n_0>n$ 
such that $\overline{\|\Delta-\overline{\Delta}\|^2}\ge \xi\varepsilon^2$,
which would overtly contradict \eq{concdeltoso}.
Because of normalisation, the following equation, complementary to the previous one, holds as well
\be
\forall \varepsilon>0 \;{\rm and}\; \forall \xi>0\, , \; \;\exists\;\tilde{n} \,|\, 
\,\forall\, n>\tilde{n} \,:\,\Gamma_{n}(D_{\varepsilon})>(1-\xi) \, .\label{concecomp}
\ee 
Together with \eq{concdeltoso}, the two previous statements entail
\be
\forall \varepsilon>0 \;{\rm and}\; \forall \xi>0\, , \; \;\exists\;\tilde{n} \,|\, 
\,\forall\, n>\tilde{n} \,:\,\int_{R_{\varepsilon}}\|\Delta\|^2\Gamma_{n}({\rm d}\Delta)<\xi \, . \label{delbridge}
\ee
The consequences of these facts for the local von Neumann entropy can be easily derived by 
exploiting the following simple properties.
As apparent from \eq{inva}, $\Delta_{d}^{m}\ge(\nu_{j}^{m})^2$, $\forall\,j$ and $\forall\,d$: 
any symplectic invariant is larger than any symplectic eigenvalue. Moreover, for the function $h(x)$ 
-- defining the von Neumann entropy according to \eq{vneu} -- one has $x^2>h(x)$ for $x\ge1$ 
(as is the case for the symplectic eigenvalues, lower bounded by $1$ because of the uncertainty principle). 
This results into $\|\Delta\|^2\ge S$. One can also show that 
$\|\Delta\|^2\ge S^2$. Therefore, bounds analogous to (\ref{delbridge}) hold for the integrals 
over $R_{\varepsilon}$ of $S$ and $S^2$ as well.
Let us also note that the function $g(\Delta)$, relating the symplectic invariants to the von Neumann entropy, 
is certainly continuous (as it relates a continuous function of the eigenvalues of a strictly positive matrix 
to the coefficients of the characteristic polynomial of the matrix).
Putting everything together we find
\be
\fl\forall\, \xi>0\, , \; \;\exists\;\tilde{n} \,\;{\rm s.t.}\;\, \forall\,n>\tilde{n} \,:\, 
(g(\overline{\Delta})-\xi)(1-\xi)+\xi \le \overline{S}\le (g(\overline{\Delta})+\xi) + \xi \, ,
\ee
which is just equivalent to \eq{avvneu}.
In the previous inequalities the continuity of $g(\Delta)$ has been invoked and the integral 
giving the average $\overline{S}$ has been decomposed into an integral over $D_{\varepsilon}$
and an integral over $R_{\varepsilon}$. An identical argument holds for the average $\overline{S^2}$,
which can be shown to converge to $g(\overline{\Delta})^2$, thus proving \eq{varvneu} as well 
and completing our treatment.

\section{Maximal entanglement for given energy}\label{maximal}

We derive here an expression for the maximal value of the entanglement (\ref{minpur})
of pure Gaussian states for given energy (under a generic $m+n$ mode bipartition), 
by adopting an explicit phase space approach. 
To begin with, let us remark that any pure Gaussian state of $m+n$ modes 
can be reduced, by local (with respect to the $m+n$ mode bipartition) 
symplectic operations, into the tensor product of $m$ two-mode squeezed states 
and of $n-m$ uncorrelated vacua (here we assume, without loss of generality, $m\le n$)
\cite{botero03,qinfc03}.
Now, the local reduction of such a state pertaining to the $m$-mode system is a Gaussian state
with CM in Williamson form. We will now prove that, amongst the CM's with the same symplectic spectrum,
the Williamson form is the one for which the second moments' contribution to the energy
$E=\tr{\sig}$ is minimal.
To this aim, let us recall that a generic CM $\sig$ with Williamson form $\gr\nu$
(and symplectic spectrum given by $\{\nu_j, \,{\rm for}\, 1\le j \le m\}$
can be written as
\be
\sig = O'^{\sf T}ZO^{\sf T} {\gr\nu} O Z O' \; ,
\ee
where the Euler decomposition of a generic symplectic transformation has been applied.
Clearly, the transformation $O'$ do not affect the energy and can be neglected in what follows.
As for the other terms, let us define
$Z'=\,{\rm diag}(z_1,\ldots,z_m)$, $\nu'=\,{\rm diag}(\nu_1,\ldots,\nu_m)$ 
and $X$ and $Y$ such that $(X+iY)\in U(m)$ [by virtue of the isomorphism of 
\eq{isu}], to obtain:
\bea
\tr{\sig} &=& \tr\left[{(Z'^2+Z'^{-2})(X^{\sf T}\nu'X+Y^{\sf T}\nu'Y)}\right] \ge \nonumber\\
&&\ge 2\tr\left[{X^{\sf T}\nu'X+Y^{\sf T}\nu'Y}\right] = 2\tr{\nu'} = \tr{\nu} \, , \nonumber
\eea
where the inequality ensues from a basic property of the trace of a product of 
positive matrices (see \cite{bathiatrace} and notice that, obviously, the eigenvalues 
of $(Z'^2+Z'^{-2})$ are larger than 2) and from the fact that the transformation 
parametrised by $X$ and $Y$ is orthogonal and preserves the trace. 
The state achieving maximal entanglement for given energy $E$ is thus 
a tensor product of $m$ two-mode squeezed states (being the state with minimal energy 
for given entanglement). 
The von Neumann entropy $S$ of the $m$-mode reduction of such a state is given 
by \eq{vneu}, 
with the local symplectic eigenvalues $\{\nu_j\}$ subject to the constraint 
$E = 4 \sum_{j=1}^{m} \nu_j + 2 (n-m)$. Because of the concavity of $h(x)$, 
the optimal choice of $\nu_j$'s, maximising the local entropy, is simply given by 
$\nu_j = \frac{E-2(n-m)}{4m} \; \forall j$, in compliace with the previous constraint.
Finally, the maximal von Neumann entropy $S_{max}(m,n,E)$ of an $m$-mode reduction of a $(m+n)$-mode
pure Gaussian state with $m\le n$ and total energy $E$ is 
\be
S_{max}(m,n,E) = m \, h\left(\frac{E-2(n-m)}{4m}\right)
\label{maxvneu} \; .
\ee
Note that this expression diverges in the thermodynamical limit.
The corresponding minimal purity is given by \eq{minpur}.\bigskip

\section*{Acknowledgments}
\noindent We thank R.~Oliveira, J.~Eisert and K. Audenaert for discussions.
This research is part of the QIP IRC www.qipirc.org (GR/S82176/01) 
and was supported by the EU Integrated Project QAP, 
the Institute for Mathematical Sciences of Imperial College London and by the Royal Society. 
A.S.~is a Marie Curie Fellow.


\begin{thebibliography}{99}
\bibitem{gennote}Even though the agreement between 
the two pairs of equations could allow for a more general choice,
i.i.d. variables are by far the most convenient assumption 
for the sake of clarity of both mathematical description 
and physical interpretation (see Ref.~\cite{nosotros}).
\end{thebibliography}
\end{document}